\documentclass[12pt]{iopart}

\usepackage{iopams}
\usepackage[english]{babel}
\usepackage{graphicx}
\usepackage{cite}
\usepackage{color}
\usepackage{bbm}

\begin{document}

\title{Optical trimer: A theoretical physics approach to waveguide couplers}

\author{A. Stoffel}
\address{Instituto Nacional de Astrof\'{\i}sica, \'Optica y Electr\'onica, Calle Luis Enrique Erro No. 1, Sta. Ma. Tonantzintla, Pue. CP 72840, M\'exico}
\author{ S. F. Caballero-Benitez} 
\address{Instituto de F\'\i sica, Universidad Nacional Aut\'onoma de M\'exico, Apartado Postal 20-364, C.P. 01000 M\'exico D. F. M\'exico.}
\author{B. M. Rodr\'\i guez-Lara$^*$}
\address{Instituto Nacional de Astrof\'{\i}sica, \'Optica y Electr\'onica, Calle Luis Enrique Erro No. 1, Sta. Ma. Tonantzintla, Pue. CP 72840, M\'exico}
\address{Photonics and Mathematical Optics Group, Tecnol\'ogico de Monterrey, Monterrey 64849, M\'exico.}
\ead{bmlara@itesm.mx}



\begin{abstract}
We study electromagnetic field propagation through an ideal, passive, triangular three-waveguide coupler using a symmetry based approach to take advantage of the underlying $SU(3)$ symmetry.
The planar version of this platform has proven valuable in photonic circuit design providing optical sampling, filtering, modulating, multiplexing, and switching.
We show that a group-theory approach can readily provide a starting point for design optimization of the triangular version.
Our analysis is presented as a practical tutorial on the use of group theory to study photonic lattices for those not familiar with abstract algebra methods.
In particular, we study the equilateral trimer to show the relation of pearl-necklace arrays with the Discrete Fourier Transform due to their cyclic group symmetry, and the isosceles trimer to show its relation with the golden ratio and its ability to provide stable output at a single waveguide.  
We also study the propagation dependent case of an equilateral trimer that linearly increases or decreases its effective coupling with the propagation distance. 
For the sake of completeness, we go beyond the standard optical-quantum analogy to show that it is possible to derive coupled-mode equations for intensity and phase, and that it is also possible to calculate envelopes for the propagation of all possible inputs belonging to an energy class, as well as individual input field amplitudes, without the need of point-to-point propagation.
\end{abstract}


\maketitle

\section{Introduction}

The planar three-waveguide coupler \cite{Iwasaki1975p100} has proven a reliable platform for optical devices. 
It has been shown to provide improving interferometers \cite{Sheem1981p3865}, tunable sampling and filtering \cite{Haus1981p2321},  modulation \cite{Donelly1985p18} and power coupling \cite{Charczenko1989p202} in voltage driven systems, as well as power dividers and combiners in passive devices \cite{Donelly1983p417,Donelly1986p610,Donelly1987p401,Kubo1989p1924} that have allowed efficient signal referencing for integrated optical biosensors \cite{Luff1998p583}.
In most of the reported literature, optimization seems the standard approach  favored by the optics community to design waveguide couplers \cite{Su1989p1666,Petrovic2015p139} but, recently, analogies with quantum mechanical systems have provided a complementary approach \cite{PerezLeija2013p012309,PerezLeija2013p022303}. 
This has also impacted the design of planar three-waveguide couplers that, for example,  have provided fast, robust directional beam coupling designed either by standard optimization \cite{Ng1999p475,Schneider2001p129,Narevicius2005p3362} or by quantum analogies \cite{Paspalakis2006p30,Salandrino2009p4524,Tseng2013p2478,RodriguezLara2014p013802}.

Here, our aim is to motivate photonic designers to go beyond the standard analogies between photonic lattices and quantum systems. 
We will try our best to bridge the gap between theoretical physics and optics to show how the underlying symmetries of a photonic lattice can shed light onto the design process. 
For this, we will use a general version of the three-waveguide coupler.
In the next section, we will introduce the mode-coupling model and expose its underlying $SU(3)$ symmetry. 
Then, we will extend previous work on Lie group design \cite{Vance1996p765} and construct a propagator for any given physical configuration using a Gilmore-Perelomov coherent state approach \cite{VillanuevaVergara2015p} using the full Wei-Norman method \cite{Wei1963p575}.
In order to provide practical examples, we will focus on arrays of identical waveguides, which we call optical trimers thereon, with three identical constant couplings and relate them to the discrete Fourier transform through the cyclic group.
We will also present results for the optical trimer with just two identical couplings, and relate them to the golden ratio and devices allowing a single intensity stable output. 
Furthermore, we will study the case of a shrinking or expanding equilateral trimer where the three identical effective couplings vary with the propagation distance.
Next, for the sake of completeness, we will show that it is possible to recover the coupled mode theory equations for field intensity and phase if we use the optical-quantum analogy in the classical limit.
We will also show that a classical mechanical analysis of the optical-quantum analogy can predict boundary limits for all possible propagation classes defined by a device, as well as propagation envelopes for individual initial inputs without the need of actual propagation.
Finally, we will present a summary and discuss possible extensions allowed by linear and nonlinear three-waveguide couplers.

\section{Triangular three-waveguide coupler}

\begin{figure}
\centering \includegraphics[scale= 1]{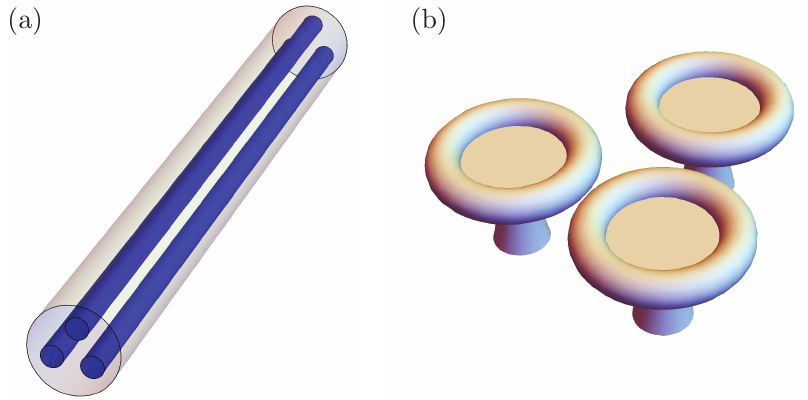}
\caption{(Color online) Some experimentally feasible realizations for triangular waveguide couplers, (a)  multicore optical fibers or inscribed photonic waveguides, (b) whispering-gallery mode cavities.}
\label{fig:Fig1}
\end{figure}

Light propagating through an ideal, triangular three-waveguide coupler can be described by coupled mode theory, c.f. \cite{RodriguezLara2015p068014} and references therein,
\begin{eqnarray} \label{eq:CMT}
-i \partial_{z} \left( \begin{array}{c} \mathcal{E}_{1}(z) \\ \mathcal{E}_{2}(z) \\  \mathcal{E}_{3}(z) \end{array} \right) =  \left( \begin{array}{ccc} 
\omega_{1}(z)  & g_{12}(z) & g_{13}(z) \\
g_{12}(z) & \omega_{2}(z) & g_{23}(z) \\
g_{13}(z) & g_{23}(z) & \omega_{3}(z)
\end{array} \right) \left( \begin{array}{c} \mathcal{E}_{1}(z) \\ \mathcal{E}_{2}(z) \\  \mathcal{E}_{3}(z) \end{array} \right).
 \nonumber \\
\end{eqnarray}
Here, the complex field amplitude at the $j$th waveguide is given by $\mathcal{E}_{j}(z)$, the real effective refractive index at the $j$th waveguide is $\omega_{j}(z)$, and the real effective coupling between the $j$th and $k$th waveguides is $g_{jk}(z)$.
These complex field equations can be cast in a Schr\"odinger-like form,
\begin{eqnarray}
- i \partial_{z} \vert \mathcal{E}(z) \rangle = \hat{H}(z) \vert \mathcal{E}(z) \rangle,\label{eq:SchLike}
\end{eqnarray}
where kets and operators in Dirac notation represent column vectors and square matrices, in that order.
We can normalize the intensity, $\sum_{j} \vert \mathcal{E}_{j}(z) \vert^2 =1$, as we are dealing with an ideal lossless device.
Experimental realization of this model include, but are not limited to, laser inscribed photonic waveguides \cite{Szameit2010p163001} and  multicore optical fibers, Fig. \ref{fig:Fig1}(a), whispering-mode cavities \cite{Peng2014p394}, Fig. \ref{fig:Fig1}(b), or microwave resonators \cite{FrancoVillafane2013p170405}.

The formal solution to this ordinary differential equation,
\begin{eqnarray}
\vert \mathcal{E}(z) \rangle = \hat{U}(z) \vert \mathcal{E}(0) \rangle,
\end{eqnarray}
is provided by an ordered exponential \cite{Magnus1954p649,Blanes2009p151}, 
\begin{eqnarray} 
\hat{U}(z) = \mathrm{Texp} \left[ \int_{0}^{z} \hat{H}(x) dx \right].
\end{eqnarray}
Usually, it is not straightforward to calculate this propagator,
but underlying symmetries simplify this endeavor \cite{Lie1880p441,Wei1963p575,Neumaier2008}.
While group theory is extensively used in mathematical optics \cite{Wolf2004,Lakshminarayanan2012}, it may be possible that the standard Lie algebra approach may look more complicated than it actually is for those outside that field. 
We hope that the following can help vanquish that feeling.

\section{Group theory approach}

Group theory, as an instrument to explore the underlying structure of mathematical models describing the physical world, brings a layer of abstraction into physics that allows deeper insight.
As such, it has become an essential tool in quantum mechanics.
Coupled mode theory delivers a Schr\"odinger-like form describing light propagating through arrays of coupled waveguides, thus, the use of group theory to calculate propagation in these systems seems like a natural step.

\begin{figure}
\centering  \includegraphics[scale= 1]{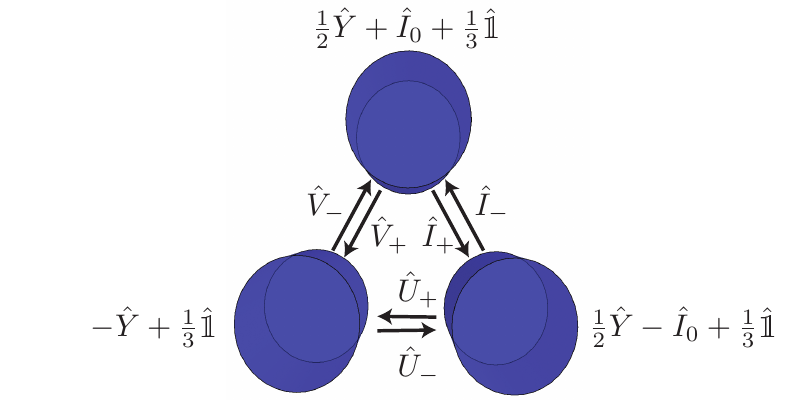}
\caption{(Color online) Schematic of relations between the $SU(3)$ matrices and the effective refractive indices and couplings in the optical trimer.}
\label{fig:Fig2}
\end{figure}

The mode-coupling matrix $\hat{H}$ for our triangular three-waveguide coupler is a unitary matrix of rank three with trace equal to $\sum_{j} \omega_{j}(z)$. 
It is useful to decompose it into a unit matrix part and a traceless part, 
\begin{eqnarray}
	\hat{H}(z) &=& \frac{1}{3} \sum_{j=1}^{3} \omega_{j}(z) \hat{\mathbbm{1}} + \hat{\mathcal{H}}(z).
\end{eqnarray}
The traceless part can be written in terms of the special unitary group $SU(3)$ which is  a household name in physics often related to the work of Gell-Mann \cite{GellMann1961} and Ne'eman \cite{Neeman1961p222},
\begin{eqnarray}
	\hat{\mathcal{H}}(z)&=& \frac{1}{2} \omega_{y}(z) \hat{Y}  + \omega_{i}(z) \hat{I}_{0}+ g_{12}(z) \left( \hat{I}_{+} + \hat{I}_{-} \right) \nonumber \\
			&&  + g_{23}(z) \left( \hat{U}_{+} + \hat{U}_{-} \right)  +   g_{13}(z) \left( \hat{V}_{+} + \hat{V}_{-} \right). \label{eq:MCMatrix}
\end{eqnarray}
Here, we have defined the auxiliary effective refractive index $\omega_{y}(z) = \left[ \omega_{1}(z) + \omega_{2}(z) - 2 \omega_{3}(z)\right]/2$, $\omega_{i}(z)= \omega_{1}(z)-\omega_{2}(z)$, and used the following representation for the $SU(3)$ group \cite{Ticciati1999}, 
\begin{eqnarray}
\hat{Y} = \frac{1}{3} \left( \begin{array}{ccc} 
1&0&0\\0&1&0\\0&0&-2    \end{array}\right), \quad
\hat{I}_{0} = \frac{1}{2} \left( \begin{array}{ccc} 1&0&0\\0&-1&0\\0&0&0 \end{array}\right), \quad  
\nonumber\\
\hat{I}_{+} = \left( \begin{array}{ccc} 0&1&0\\0&0&0\\0&0&0 \end{array}\right), \quad 
\hat{I}_{-} = \left( \begin{array}{ccc} 0&0&0\\1&0&0\\0&0&0 \end{array}\right), \quad \nonumber\\
\hat{U}_{+} = \left( \begin{array}{ccc} 0&0&0\\0&0&1\\0&0&0 \end{array}\right), \quad 
\hat{U}_{-} = \left( \begin{array}{ccc} 0&0&0\\0&0&0\\0&1&0 \end{array}\right), \quad \nonumber \\
\hat{V}_{+} = \left( \begin{array}{ccc} 0&0&1\\0&0&0\\0&0&0 \end{array}\right), \quad 
\hat{V}_{-} = \left( \begin{array}{ccc} 0&0&0\\0&0&0\\1&0&0 \end{array}\right). \label{eq:gens}
\end{eqnarray}
It is straightforward to relate matrices $\hat{I}_{\pm}$ to the couplings between the first and second waveguides, matrices $\hat{U}_{\pm}$ to the couplings between second and third waveguides, and matrices $\hat{V}_{\pm}$ to the couplings between third and first waveguides.
A little bit of algebra allows us to see that the sum of matrices $\hat{Y}/2 + \hat{I}_{0} + \hat{\mathbbm{1}}/3$ is related to the effective refractive index of the first waveguide, matrix $\hat{Y}/2 - \hat{I}_{0} + \hat{\mathbbm{1}}/3$ to the effective refractive index of the second waveguide, and matrix $-\hat{Y} + \hat{ \mathbbm{1}}/3$ to that of the third waveguide, as shown schematically in Fig. \ref{fig:Fig2}.

At this point, our original Schr\"odinger-like equation is written in terms of the identity matrix and a linear combination of Lie group generators for $SU(3)$.
The identity part only induces and overall phase,
\begin{eqnarray}
e^{i \phi(z) \hat{\mathbbm{1}}} = e^{\frac{i}{3} \int_{0}^{z} \left(\omega_{1}(\zeta) + \omega_{2}(\zeta) + \omega_{3}(\zeta) \right) d\zeta } ~\hat{\mathbbm{1}} .
\end{eqnarray}
such that the propagator can be rewritten,
\begin{eqnarray}
	\hat{U}(z) =  e^{ \frac{i}{3} \int_{0}^{z} ( \omega_{1}(\zeta) + \omega_{2}(\zeta) + \omega_{3}(\zeta) ) d\zeta} ~\hat{\mathcal{U}}(z).
\end{eqnarray}
Now, for the $SU(3)$ part, $\hat{\mathcal{U}}(z)$, Wei and Norman demonstrated that any such equation can be treated by an algebraic method providing the following propagator \cite{Wei1963p575},
\begin{eqnarray}
	\hat{\mathcal{U}}(z) = \prod_{j=1}^{8} e^{i \theta_{j}(z) \hat{X}_{j}},
\end{eqnarray}
where the $su(3)$ algebra elements, $e^{i \theta_{j}(z) \hat{X}_{j}}$, are just the exponential map of the $SU(3)$ group generators, $ \hat{X}_{j} \in  \left\{ \hat{Y}, \hat{I}_{0}, \hat{I}_{\pm}, \hat{U}_{\pm}, \hat{V}_{\pm} \right\}$, and the functions $\theta_{j}(z)$ are complex functions ruled by the dynamics provided by the mode-coupling matrix, $\hat{\mathcal{H}}(z)$.
There is no apriori ordering of $su(3)$ elements to write the propagator, however, the values of the $\theta_{j}(z)$ functions do depend on the chosen order and different orderings have been studied in the quantum optics literature \cite{Dattoli1987p1582,Dattoli1991p1247,Dattoli1990p236,Gnutzmann1998p9871}.
We will choose a particular ordering,
\begin{eqnarray}
\hat{\mathcal{U}}(z) &=& e^{i \iota_{+}(z) \hat{I}_{+}} e^{i \mu_{+}(z) \hat{U}_{+}}  
e^{i \nu_{+}(z) \hat{V}_{+}} e^{ \iota_{0}(z) \hat{I}_{0}} \nonumber \\ 
&& \times e^{i y_{0}(z) \hat{Y}}  e^{i \nu_{-}(z) \hat{V}_{-}} e^{i \mu_{-}(z) \hat{U}_{-}} e^{i \iota_{-}(z) \hat{I}_{-}}, \label{eq:prop}
\end{eqnarray}
that keeps us in line with the idea of understanding propagation through waveguide lattices as generalized Gilmore-Perelomov coherent states \cite{VillanuevaVergara2015p}.

The next step is straightforward but cumbersome, we substitute the formal solution, $\vert \mathcal{E}(z) \rangle$, using the propagator above and being careful in keeping the ordering through the derivation process. 
Then, we use the actions of elements of the $su(3)$ algebra on elements of the $SU(3)$ group \cite{Nelson1967p857} to find the differential equation set for the auxiliary functions, this reduces the problem to solving a so-called triangular system where we first need to solve a complex matrix Riccati equation,
\begin{eqnarray}\label{eq:fstRiccati}
\left( \begin{array}{c} \iota_{+}^{\prime} \\ \nonumber
\nu_{+}^{\prime}  \end{array}\right) &=& 
\left(\begin{array}
	{c} g_{12} \\ 
	g_{13}  
\end{array}\right) \\\nonumber
&&+ i 
\left(\begin{array}{cc}  
 	\omega_{i} & - g_{23} \\
 	- g_{23} & \frac{1}{2} 
 	\left[ 2\omega_{y} +
 	 \omega_{i}\right]
\end{array}\right)
\left( \begin{array}{c} 
 	\iota_{+} 
 	\\ \nu_{+}
\end{array}\right)
\\
&&+ \left( \begin{array}{c} 
\iota_{+} 
\\ \nu_{+}
\end{array}\right)
\left(\begin{array}{c}  
g_{12} \\
g_{13}
\end{array}\right)^T
\left( \begin{array}{c} 
\iota_{+} 
\\ \nu_{+}
\end{array}\right)
\end{eqnarray}
then by direct substitution we can get to solve another Riccati equation,
\begin{eqnarray}\label{eq:sndRiccati}
\mu_{+}^{\prime} &=& \left(g_{23} + i g_{13} \iota_{+} \right)\mu_{+}^{2} + \Big[ g_{13} \nu_{+} - g_{12} \iota_{+}  \nonumber \\
 &&  + \frac{i}{2} \left(2\omega_{y} - \omega_{i} \right)\Big] \mu_{+}  + i g_{12} \nu_{+} + g_{23}.
\end{eqnarray}
Once these are solved, the rest can be found by direct integration,
\begin{eqnarray}
\iota_{0}^{\prime} &=& \omega_{i} - i 2  g_{12} \iota_{+} + i g_{23} \mu_{+} - g_{13} \left( \iota_{+} \mu_{+} + i \nu_{+} \right) ,\label{eq:diff1} \\
y_{0}^{\prime} &=& \omega_{y} - i\frac{3}{2}  
[ g_{13}   \nu_{+} +    
( g_{23} + i  g_{13} \iota_{+})\mu_{+}], \label{eq:diff2} \\
\nu_{-}^{\prime} &=& g_{13} e^{i  \left( y_0 + \frac{1}{2} \iota_0 \right)}
-  e^{i\iota_{0}}   \left( g_{13} \mu_{+} + i  g_{12} \right) \mu_{-} , \label{eq:mu:min} \\
\mu_{-}^{\prime} &=&  e^{i  \left( y_0 - \frac{1}{2} \iota_0 \right)}
( g_{23} + i g_{13} \iota_{+}),\\
\iota_{-}^{\prime} &=& e^{i \iota_{0}} \left( g_{12} - i g_{13}\mu_{+} \right),\label{eq:iota:min}
\end{eqnarray}
where, for the sake of space, we have used $f \equiv f(z)$ and $f' \equiv \partial_{z} f(z)$ for auxiliary functions and couplings.

Non-linear differential equations are known to be hard to solve and finding a solution 
often requires intuition and knowledge of the system being analyzed. 
Before delving into details, we would like to point out a key feature of passive, lossless optical models that we are analyzing, their mode-coupling matrices are real symmetric, $\hat{H}^{T}(z) = \hat{H}(z)$ where the operation $O^{T}$ stands for transposition,  and, as a direct consequence, the propagator fulfills $\hat{U}^{T}(z) \hat{U}(z) = \hat{\mathbbm{1}}$.
This feature allows us to conclude that the propagator functions are symmetric,
\begin{eqnarray}\label{eq:symmCon}
	\xi_{+}(z)&=&\xi_{-}(z), \quad \xi = \iota, \nu, \mu.
\end{eqnarray}
Most of the time, numerical methods would be the only way to find these auxiliary functions but standard and matrix Riccati equations are well known in the field of mathematical physics and control theory and we can rely on known solutions from those areas.

An alternate approach can be followed if we write the propagator as a matrix,
\begin{eqnarray}
\hat{\mathcal{U}}(z) = \left( \begin{array}{ccc} 
\Xi(z) & \Sigma(z) & \Theta(z) \\
\Sigma(z) & \Pi(z)    & \Delta(z) \\
\Theta(z) & \Delta(z) & \Gamma(z)
\end{array}  \right),
\end{eqnarray}
where, for reasons that will become apparent in a moment, we introduced a set of five auxiliary functions,
\begin{eqnarray}
	\Gamma(z) &=& e^{- i \frac{2}{3}  y_{0}(z)}, \\
	\Delta(z)&=& i \Gamma(z) \mu_{+}(z), \\
	\Theta(z)&=& \Gamma(z) \left[ -\iota_{+}(z) \mu_{+}(z) + i \nu_{+}(z) \right], \\
	\Pi(z)&=& \Gamma(z) \left[ e^{i y_0 (z)}e^{-i \frac{1}{2}\iota_0 (z)}
		-\mu_{+} (z)^2 \right], \\	
	\Sigma(z)&=& i \iota_{+}(z) \Pi(z) - \Gamma(z) \mu_{+} (z) \nu_{+}(z)), 
\end{eqnarray}
and the sixth can be written in terms of all others,  
\begin{eqnarray}
\Xi(z) = \frac{1 + \Pi(z) \Theta^{2}(z) + \Gamma(z) \Sigma^{2}(z)- 2 \Delta(z) \Theta(z) \Sigma(z)}{ \Pi(z) \Gamma(z) - \Delta^{2}(z)}. \nonumber \\
\end{eqnarray}
The original and auxiliary functions are connected,
\begin{eqnarray}
\iota_{+}(z) &=& i \frac{\Gamma(z)\Sigma(z) - \Delta(z)\Theta(z)}{\Delta^2(z)-\Gamma(z)\Pi(z)}, \\
\mu_{+}(z) &=& -i \frac{\Delta(z)}{\Gamma(z)},\\
\nu_{+}(z) &=& i \frac{\Pi(z)\Theta(z) - \Delta(z)\Sigma(z)}{\Delta^2(z)-\Gamma(z)\Pi(z)} ,\\
\iota_{0}(z) &=& i 2 \log \frac{\Gamma(z) \Pi(z) - \Delta^{2}(z)}{\Gamma^{\frac{1}{2}}(z) } , \\
y_{0}(z) &=& i \frac{3}{2} \log \Gamma(z) .
\end{eqnarray}
Note that the phase functions $y_0(z)$ and $i_0(z)$ are of logarithmic nature and the others are quotients of the products of the solution basis. 
We now have the simplest matrix differential equation for the propagator \cite{Reid1939p414,Levin1959p519}, 
\begin{eqnarray}
\partial_{z} \hat{\mathcal{U}}(z) = i \mathcal{H} \hat{\mathcal{U}}(z) ,
\end{eqnarray}
with the initial condition $\hat{\mathcal{U}}(0) = \hat{\mathbbm{1}}$, that translates into,
\begin{eqnarray}
\Theta(z)= 0 , \;\Delta(z) =  0, \; \Sigma(z)= 0, \;
\Gamma(z)= 1, \; \Pi(z)= 1. \nonumber \\ \label{eq:initsb}
\end{eqnarray}
This differential equation system is overdetermined due to the characteristics of the original matrix and provides the following identities,
\begin{eqnarray}
g_{12} \left( \Pi - \Xi \right) + g_{13} \Delta - g_{23} \Theta + \omega_{i} \Sigma &=& 0,  \nonumber \\\\
g_{12} \Delta + g_{13} \left( \Gamma - \Xi \right) - g_{23} \Sigma + \left(  \omega_{y} +\frac{1}{2}\omega_{i}   \right) \Theta &=& 0, \nonumber \\ \\
g_{12} \Theta - g_{13} \Sigma(z) + g_{23} \left( \Gamma - \Pi \right)  + \left(  \omega_{y} - \frac{1}{2}\omega_{i}\right) \Delta &=& 0.  \nonumber \\
\end{eqnarray}
Again, for waveguide couplers that depend on the propagation distance, this formal solution will be analytically tractable only for a handful of cases, for example the planar three-waveguide coupler \cite{RodriguezLara2014p013802}. 

\subsection{Constant optical trimer}
While we have provided a formal solution to propagation through the most general triangular waveguide coupler, considering a particular solution may help build further intuition. 
For the sake of simplicity, let us now consider the constant optical trimer, which is a triangular three-waveguide array with constant couplings and identical waveguides. 
We will introduce the dimensionless propagation parameter $\zeta = g_{12} z$, such that the mode-coupling differential equation becomes 
\begin{eqnarray}
- i \partial_{\zeta} \vert \mathcal{E}(\zeta) \rangle = \hat{\mathcal{H}} \vert \mathcal{E}(\zeta) \rangle, 
\end{eqnarray}
with the mode-coupling matrix,
\begin{eqnarray}
\hat{\mathcal{H}} = \left( \begin{array}{ccc} 
0  & 1 & \alpha  \\
1 & 0 & \beta \\
\alpha & \beta & 0
\end{array} \right), \label{eq:hmlt2}
\end{eqnarray}
given in terms of the dimensionless parameters,
\begin{eqnarray}
\alpha =\frac{g_{13}}{g_{12}}, \quad \beta=\frac{g_{23}}{g_{12}}.
\end{eqnarray}
Now, we can use the results above to build a particular solution, but it is well known that a set of linear first order differential equations is equivalent to a single linear differential equation of higher order.
After some algebra, we can derive a third order differential equation for $\Delta(\zeta)$,
\begin{equation}
\Delta^{\prime\prime\prime}(\zeta) + i(1+\alpha^2+\beta^2)\Delta ^{\prime}(\zeta) -2 \alpha\beta\Delta(\zeta) = 0,\label{eq:ddiff}
\end{equation}
with boundary conditions, 
\begin{eqnarray}
\Delta(0) = 0, \quad \Delta^{\prime}(0) = i\beta, \quad \Delta^{\prime\prime}(0) = -\alpha.
\end{eqnarray}
Note that the remaining auxiliary functions are straightforward to calculate, 
\begin{eqnarray}
\Theta(\zeta)&=& \frac{\beta(1+\beta^2)\Delta(\zeta) - i \alpha\Delta^{\prime}(\zeta)+\beta\Delta^{\prime\prime}(\zeta)}
	{\alpha(1-\beta^2)}  \\
\Gamma(\zeta)&=& \frac{-(1+\beta^2)\Delta(\zeta) +i \alpha\beta\Delta^{\prime}(\zeta)-\Delta^{\prime\prime}(\zeta)}
	{\alpha(1-\beta^2)} \\
\Sigma(\zeta)&=& \frac{\beta(\alpha^2+\beta^2)\Delta(\zeta) -i \alpha\Delta^{\prime}(\zeta)+\beta\Delta^{\prime\prime}(\zeta)}
	{\alpha^2-\beta^2}  \\
\Pi(\zeta)&=& \frac{-\alpha(\alpha^2+\beta^2)\Delta(\zeta) +i \beta\Delta^{\prime}(\zeta)-\alpha\Delta^{\prime\prime}(\zeta)}
	{\alpha^2-\beta^2}.
\end{eqnarray} 

\begin{figure}[htbp]
\centering \includegraphics[scale= 1]{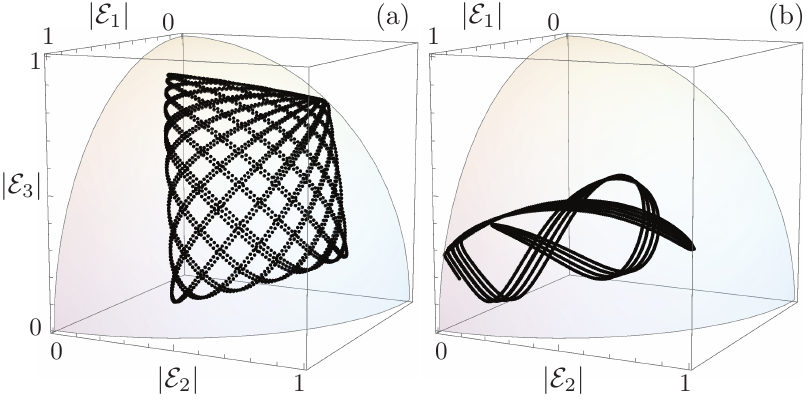}
\caption{(Color online) Propagation trajectories of absolute amplitudes, $(\vert \mathcal{E}_{1}(z) \vert, \vert \mathcal{E}_{2}(z) \vert, \vert \mathcal{E}_{3}(z) \vert )$, for two constant optical trimers with random parameters and random normalized initial field distributions.}
\label{fig: Fig3}
\end{figure}

The main auxiliary function $\Delta(\zeta)$ has the following solution,
\begin{eqnarray}
\Delta(\zeta) = \delta_{1} \: e^{i \gamma_1 \zeta} + \delta_{2} \: e^{i \gamma_2 \zeta} +  \delta_{3} \: e^{i \gamma_3 \zeta},
\end{eqnarray}  
where the constant parameters  $\gamma_{j}$ are the eigenvalues of the mode-coupling matrix determined by the characteristic polynomial, a reduced cubic,
\begin{eqnarray}
\gamma_{j}^{3} - (1+ \alpha^{2} + \beta^{2}) \gamma_{j} - 2 \alpha \beta = 0. \label{eq:poly}
\end{eqnarray}
It is straightforward to notice that there are three different real eigenvalues for real, positive, non-zero coupling parameters, $\alpha, \beta > 0$.
These proper values can be writen in a closed but non-compact form, so we will not write them explicitly.
Furthermore, the coefficients are given by
\begin{eqnarray}
\delta_{1} &=& \frac{\alpha-\beta(\gamma_2+\gamma_3)}{(\gamma_1-\gamma_2)(\gamma_1-\gamma_3)},\\
\delta_{2} &=& \frac{\alpha-\beta(\gamma_1+\gamma_3)}{(\gamma_2-\gamma_1)(\gamma_2-\gamma_3)},\\
\delta_{3} &=& \frac{\alpha-\beta(\gamma_1+\gamma_2)}{(\gamma_3-\gamma_1)(\gamma_3-\gamma_2)}. 
\end{eqnarray}  
Thus, the propagator functions, $\iota_{\pm}(z)$, $\mu_{\pm}(z)$, $\nu_{\pm}(z)$, $\iota_{0}(z)$ and $y_{0}(z)$, will effectively contain terms involving the three eigenvalues as well as sums and differences thereof.
Figure \ref{fig: Fig3} shows the trajectories described by the absolute value of the field amplitudes, $\vert \mathcal{E}_{j}(z)\vert$, as they propagate through the constant optical trimer with random parameters and initial field distribution calculated provided by the group theory approach.

\subsection{Propagation dependent trimer with identical couplings}
The discussion above focus on constant coupling terms, ie. terms that do not exhibit z-dependence.
To unveil and demonstrate the true vigor of the Wei-Norman approach we will, in the remainder of this section, turn to a waveguide system with equal, z-dependent coupling terms and identical refractive indices.
The effective mode coupling matrix for such a system is written as:
\begin{eqnarray}\label{eq:modCppl:zDep}
\hat{\mathcal{H}}(z) =
\left( \begin{array}{ccc} 
0  & g(z) & g(z) \\
g(z) & 0 & g(z) \\
g(z) & g(z) & 0
\end{array} \right).
\end{eqnarray}
The complex matrix Riccati equation, Eq.(\ref{eq:fstRiccati}), in this particular case is symmetric, suggesting that, taking initial conditions into account, the solution also is symmetric:
\begin{eqnarray}\label{eq:symm:nuiota}	
\nu(z)=\iota(z), 
\end{eqnarray}
where we have dropped the subscript $\pm$ for the sake of simplicity as a result of  Eq.(\ref{eq:symmCon}). 
Consequently, the complex matrix Riccati differential equation, Eq.(\ref{eq:fstRiccati}) trims back to a single, complex Riccati differential equation:
\begin{eqnarray}\label{eq:redRiccati}
\iota^{\prime}(z) = g(z)[1 - i\iota(z) +2 \iota^2(z)].
\end{eqnarray}
Then, the other Riccati equation we need to solve, Eq.(\ref{eq:sndRiccati}), reads,
\begin{eqnarray}\label{eq:sndRiccati:zDep}
\mu^{\prime}(z) = 
g(z) [1+i\iota(z) + (1+i\iota(z))\mu(z)^2].
\end{eqnarray}
In light of the symmetries, we have two remaining functions, namely $y_0(z)$ and $\iota_0(z)$, 
to solve until we have a closed form complete solution that yields the propagator, Eq(\ref{eq:prop}). 
It is not difficult to see that Eq.(\ref{eq:mu:min})
to Eq.(\ref{eq:iota:min}), combined, can yield 
functions $y_0(z)$ and $\iota_0(z)$. 
It is important to note that care must be taken as the exponentials in Eq.(\ref{eq:mu:min}) to Eq.(\ref{eq:iota:min}) imply logarithms when solved for $y_0(z)$ and $\iota_0(z)$. 
Thus, the  discontinuity, or branch cut, along the negative real axis of the complex logarithm insinuate the need of additional terms, 
\begin{eqnarray}\label{eq:sndlastEq}
y_0(z)=\frac{-i}{2} \log\left[\frac{\iota^\prime(z) \mu^\prime(z)(1+i\mu(z))}{g(z)^2(1+i\iota(z))}\right] +2 \pi n_1(z), \nonumber \\ 
\end{eqnarray}
and
\begin{eqnarray}\label{eq:lastEq}
\iota_0(z)=\frac{-i}{2} \log\left[\frac{\iota^\prime(z)}{ \mu^\prime(z)}(1+i\mu(z))(1+i\iota(z))\right]+2 \pi n_2(z),  \nonumber \\
\end{eqnarray}
to counter the logarithmic discontinuities. 
Here the auxiliary functions $n_1(z)$ and $n_2(z)$ are z-dependent, integer functions such that logarithmic discontinuities will cancel out.
For the sake of completeness we state the propagator function,
given that Eq.(\ref{eq:symm:nuiota}) and Eq.(\ref{eq:symmCon}) hold true,
\begin{eqnarray}
\hat{\mathcal{U}}(z) &=& e^{i \iota(z) \hat{I}_{+}} e^{i \mu(z) \hat{U}_{+}}  
e^{i \iota(z) \hat{V}_{+}} e^{ \iota_{0}(z) \hat{I}_{0}} \nonumber \\ 
&& \times e^{i y_{0}(z) \hat{Y}}  e^{i \iota(z) \hat{V}_{-}} e^{i \mu(z) \hat{U}_{-}} e^{i \iota(z) \hat{I}_{-}}.
\end{eqnarray}


\section{Applications}

As we just saw, the constant optical trimer and the propagation dependent equilateral optical trimer are simple enough to allow us the construction of a closed form solution and, to our advantage, both are experimentally feasible.
Now the obvious question is if there is a use for them.
In the following, we will show that a judicious choice of coupling parameters provides  different types of well-defined propagation of field amplitudes that can be used for the design of integrated photonic circuits.

\subsection{Identical couplings and the discrete Fourier transform}

The mode-coupling matrix for three-identical waveguides distributed in an equilateral triangle configuration, $\alpha = \beta = 1$, 
\begin{equation}
\hat{\mathcal{H}}=\left( \begin{array}{ccc}
0 & 1 & 1 \\
1 & 0 & 1 \\
1 & 1 & 0 \end{array} \right),	 
\end{equation}
is related to the cyclic group in dimension three, 
\begin{eqnarray}
\hat{\mathcal{H}} =  \hat{Z}_{3} + \hat{Z}_{3}^{2} ,
\end{eqnarray}
where the generator of the cyclic group are the following, 
\begin{eqnarray}
\hat{Z}_{3} &=& \hat{I}_{+} + \hat{U}_{+} + \hat{V}_{-}, \nonumber \\
&=&\left(
\begin{array}{ccc}
 0 & 1 & 0 \\
 0 & 0 & 1 \\
 1 & 0 & 0 \\
\end{array}\right).
\end{eqnarray}
It is well known that the cyclic group is diagonalized, 
\begin{eqnarray}
\hat{\Lambda} = \hat{F}_{n} \hat{Z}_{n} \hat{F}_{n}^{\dagger},
\end{eqnarray}
by the discrete Fourier transform of rank $n$ given, for $n=3$,
\begin{eqnarray}
\hat{F}_{3} &=& 
\frac{1}{\sqrt{3}}
\left(
\begin{array}{ccc}
 1 & 1 & 1 \\
 1 & e^{\frac{2 i \pi}{3}} & e^{-\frac{2 i \pi}{3}} \\
 1 & e^{-\frac{2 i \pi}{3}} & e^{\frac{2 i \pi}{3}} \\
\end{array}\right),
\end{eqnarray}
and $\hat{\Lambda}$ is a diagonal rank $n$ matrix showing the $n$th-roots of the unit,
\begin{eqnarray}
\hat{\Lambda}_{mn} = \delta_{m,n} e^{ i \frac{2 \pi}{n} i}, \quad  m,n = 1,2,3,
\end{eqnarray} 
on the main diagonal.
In this particular case, it is possible to compose a propagator,
\begin{eqnarray}
U(\zeta) &=& \hat{F}_{3}^{\dagger} e^{i \hat{\Lambda}_{3} \zeta} e^{i \hat{\Lambda}_{3}^{2} \zeta} \hat{F}_{3}, \nonumber \\
&=& \frac{1}{3}\left(
\begin{array}{ccc}
 2+e^{3 i \zeta} & -1+e^{3 i \zeta} & -1+e^{3 i \zeta} \\
 -1+e^{3 i \zeta} & 2+e^{3 i \zeta} & -1+e^{3 i \zeta} \\
 -1+e^{3 i \zeta} & -1+e^{3 i \zeta} & 2+e^{3 i \zeta} \\
\end{array}
\right) e^{-i \zeta}, \nonumber \\
\end{eqnarray}
where we have used the fact that the elements of the cyclic group of rank $3$ commute between them, $\left[ \hat{Z}_{3} , \hat{Z}_{3}^{2} \right] = 0$ because $\hat{Z}_{3}^{3} = \hat{\mathbbm{1}}_{3}$.  

\begin{figure}[htbp]
\centering \includegraphics[scale= 1]{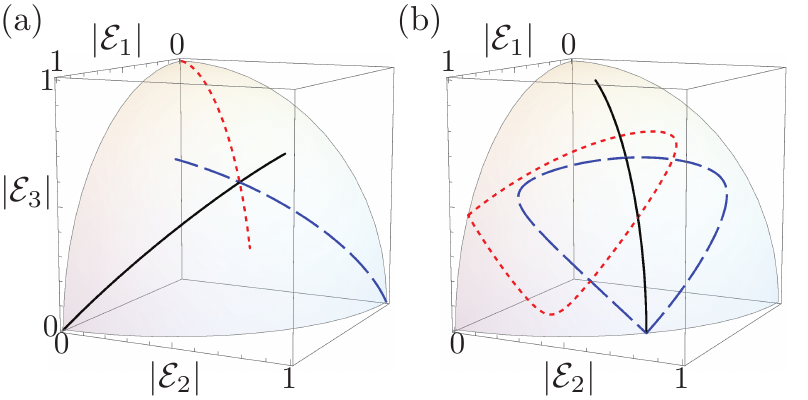}
\caption{(Color online) Propagation trajectories of absolute amplitudes, $(\vert \mathcal{E}_{1}(z) \vert, \vert \mathcal{E}_{2}(z) \vert, \vert \mathcal{E}_{3}(z) \vert )$, for initial fields impinging (a) only the zeroth (solid black), $(1,0,0)$, first (dashed blue), $(0,1,0)$, and second (dotted red), $(0,0,1)$, waveguides and (b) initial fields impinging two waveguides at a time with and without a relative phase, $(1,1,0)/ \sqrt{2}$ (solid black), $(1,i,0)/\sqrt{2}$ (dashed blue), and $(2,0,i)/\sqrt{5}$ (dotted red).}
\label{fig: Fig4}
\end{figure}

Figure \ref{fig: Fig4} shows the trajectories described by the absolute value of the field amplitudes, $\vert \mathcal{E}_{j}(z)\vert$, as they propagate. 
All trajectories will lie on the surface of an octant of the sphere due to unitary propagation, $\sum_{j} \vert \mathcal{E}_{j}(z) \vert^2 =1$.
Figure \ref{fig: Fig4}(a) shows the response to impulse input, $ \mathcal{E}_{k}(0)  = \delta_{j,k}$ with $j=1,2, 3$ and a fixed $k=1,2,3$. 
Figure \ref{fig: Fig3}(b) shows the trajectories given by initial field superpositions of the more general form: $\mathcal{E}_{j} = \alpha_{j} e^{i \phi t}$ with $\alpha \in \mathbb{R}$ and $\sum_{j} \vert \alpha_{j} \vert^{2} =1$.
Note that the roots of the mode-coupling matrix, $\hat{\mathcal{H}}$, are doubly degenerate and commesurate, 
\begin{eqnarray}
\gamma_{1,2} = -1, \quad \gamma_{3} = 2.
\end{eqnarray}
Thus, propagation involves just two commensurate eigenvalues and all the possible field amplitude propagation trajectories, $\left( \vert \mathcal{E}_{1} \vert, \vert \mathcal{E}_{2} \vert, \vert \mathcal{E}_{3} \vert  \right)$, will be closed and well defined as those shown in Fig.\ref{fig: Fig4}.

\subsection{Two identical couplings and the golden ratio}
In the case of two equal coupling parameters the unitless mode-coupling matrix becomes,
\begin{equation}
\hat{\mathcal{H}}=\left( \begin{array}{ccc}
0 & 1 & \alpha \\
1 & 0 & \alpha \\
\alpha & \alpha & 0 \end{array} \right)	 .
\end{equation}
Note that exchanging the first and second waveguides returns the same mode-coupling matrix; in other words, it is $Z_2$-invariant. 
This symmetry is reflected by the eigenvalues,
\begin{eqnarray}
\gamma_{1} = -1, \quad \gamma_{2} = \bar{\varphi}, \quad \gamma_{3} = \varphi,
\end{eqnarray}  
with
\begin{eqnarray}
\bar{\varphi} &=& \frac{1}{2} \left(1-\sqrt{8 \alpha ^2+1}\right),\\
\varphi &=& \frac{1}{2} \left( 1 + \sqrt{8 \alpha ^2+1} \right).
\end{eqnarray}
It is simply to see that $\bar{\varphi} + \varphi = 1$ and that the latter eigenvalue 
becomes the golden ratio for $\alpha = \sqrt{1/2}$.
Again, due to the $Z_{2}$-symmetry the propagator can be calculated directly using the group theory approach,
\begin{eqnarray}
\Delta(\zeta) &=& \Theta(z), \\
&=& \frac{\alpha}{\varphi - \tilde{\varphi}} \left(  e^{i \zeta \varphi }  -  e^{i \zeta \bar{\varphi} } \right), \\
\Gamma(\zeta) &=& \frac{1}{\varphi - \tilde{\varphi}} \left( \varphi e^{i \zeta \bar{\varphi} } - \bar{\varphi}  e^{i \zeta \varphi } \right), \\
\Sigma(\zeta) &=& \frac{1}{2 \left( \varphi - \tilde{\varphi} \right)} \left( \varphi e^{i \zeta \varphi } - \bar{\varphi} e^{i \zeta \bar{\varphi} } \right) - \frac{1}{2} e^{-i \zeta}, \\
\Pi(\zeta) &=& \Xi(\zeta), \\
&=& \frac{1}{2 \left( \varphi - \tilde{\varphi} \right)} \left( \varphi e^{i \zeta \varphi } - \bar{\varphi} e^{i \zeta \bar{\varphi} } \right) + \frac{1}{2} e^{-i \zeta}.
\end{eqnarray}

\begin{figure}[htbp]
	\centering \includegraphics[scale= 1]{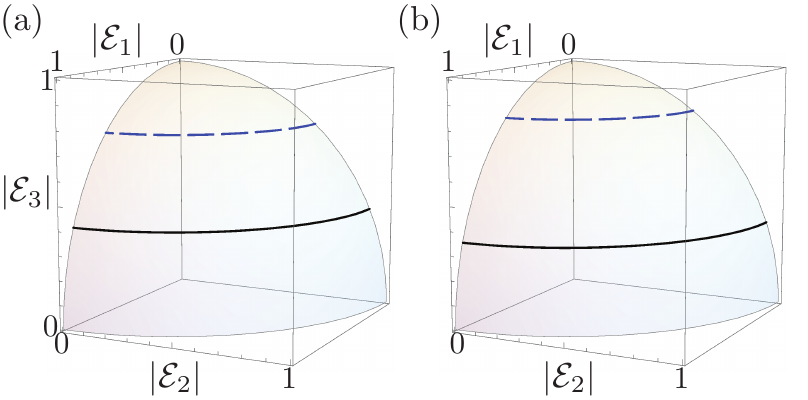}
	\caption{(Color online) Propagation trajectories of absolute amplitudes, $(\vert \mathcal{E}_{1}(z) \vert, \vert \mathcal{E}_{2}(z) \vert, \vert \mathcal{E}_{3}(z) \vert )$,  for the states with stationary field amplitude in the second waveguide related to $\varphi$ (solid black) and $\bar{\varphi}$ (dashed blue) described by Eq.(\ref{eq:StatAmp}) with (a) $\alpha = 1 / \sqrt{2}$ and (b) $\alpha = 1 / 2$ .}
	\label{fig:Fig5}
\end{figure}

Obviously, when the eigenvalues are commensurate, the field amplitude propagation trajectories will be closed and well defined. 
However, this particular symmetry might allow further interesting states.
Note that the auxiliary functions $\Delta(\zeta)$ and $\Gamma(\zeta)$ depend on only two of the three eigenfrequencies, $\varphi$ and $\bar{\varphi}$. 
Furthermore, a little bit of algebra considering an initial state without any field impinging at the zeroth waveguide, $\mathcal{E}_{1}(0)=0$, reveals that propagation in the second waveguide, $\mathcal{E}_{3}(z)$, only depends on those two auxiliary functions, $\Delta(\zeta)$ and $\Gamma(\zeta)$,
\begin{equation}
\mathcal{E}_{3}(\zeta) = \Delta(\zeta)\mathcal{E}_{2}(0)
+\Gamma(\zeta)\mathcal{E}_{3}(0).
\end{equation}
Now, we can hope that choosing a proper input will balance out one of the two 
eigenvalues that appear in the evolution of the field amplitude at the second waveguide.
That is, if the evolution only depends on a single eigenvalue, it will only change in phase and the field amplitude at the second waveguide will remain immutable. 
This is fulfilled when
\begin{eqnarray}
\mathcal{E}_{2}(0)= \frac{x}{\alpha} \mathcal{E}_{3}(0), \quad x= \varphi, \bar{\varphi}, \label{eq:IsoIniCon}
\end{eqnarray}
are met as initial conditions.
Without loss of generality, we can give the propagated amplitude vector for the two possible initial vectors defined by $\varphi$ and $\bar{\varphi}$,
\begin{equation} \label{eq:StatAmp}
\vert \mathcal{E}_{x}(\zeta) \rangle = \frac{1}{\sqrt{\mathcal{N}}}\left( \begin{array}{c}
\frac{x}{\alpha} \Sigma(\zeta) + \Delta(\zeta) \\
\frac{x}{\alpha} \Pi(z) + \Delta(\zeta) \\
e^{i x \zeta }
\end{array} \right)	\quad x= \varphi, \bar{\varphi},
\end{equation}
up to a normalization constant $\mathcal{N}$.
Figure \ref{fig:Fig5} shows propagation trajectories for the absolute field amplitudes in an isoceles constant optical trimer with parameter $\alpha = 1 /\sqrt{2}$ chosen to provide the golden ratio, Fig. \ref{fig:Fig5}(a), and  $\alpha = 1/ 2$, Fig. \ref{fig:Fig5}(b), with initial field amplitudes provided by Eq.(\ref{eq:IsoIniCon}) ensuring a stable output at the second waveguide.

\subsection{A shrinking or expanding equilateral optical trimer}

\begin{figure}[htbp]
	\centering \includegraphics[scale= 1]{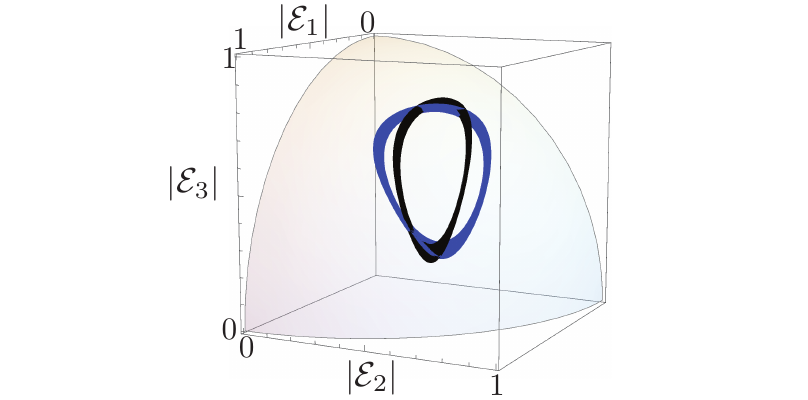}
	\caption{(Color online) Propagation trajectories of absolute amplitudes, $(\vert \mathcal{E}_{1}(z) \vert, \vert \mathcal{E}_{2}(z) \vert, \vert \mathcal{E}_{3}(z) \vert )$, for initial random fields impinging an equilateral triangular three-waveguide coupler that is shrinking with an effective coupling parameter $g(z) = 1 +0.5 z$.}
	\label{fig:Fig6}
\end{figure}

Let us turn now to a specific propagation dependent example and assume an effective linear coupling between the three waveguides, 
\begin{eqnarray}
g(z) = A + B z,
\end{eqnarray}
such that the new parameters $A$ and $B$ are real valued constants.
This type of function implies a three-waveguide coupler where the waveguide separations are symmetrically increasing or decreasing. 
Assuming this linear effective coupling, the Riccati equations, Eq.(\ref{eq:redRiccati}) and  Eq.(\ref{eq:sndRiccati:zDep}), can be solved and yield the following closed forms,
\begin{eqnarray}
\iota(z)=& i\left[1-\frac{3 C(z)}{2 C(z) + 1}\right] \\
\mu(z)=& i \tanh \left\{ \frac{1}{2} \left(\log 3 - \log
\left[ 1 + 2 C(z)\right] \right) \right\}
\end{eqnarray}
where,
\begin{eqnarray}
C(z) = e^{\frac{3}{2} i z (2 A+B z)}
\end{eqnarray}
It is straightforward to calculate their derivatives and substitute them in the remaining differential equations, Eq.(\ref{eq:sndlastEq}) and Eq.(\ref{eq:lastEq}), and solve them by direct integration,
\begin{eqnarray}
y_0(z)=
-\frac{1}{2} i \log \left\{  \frac{27 ~ C(z)}{\left[ 2+ C(z)\right]^3} \right\} + \frac{1}{2}N(z)
\end{eqnarray}
and 
\begin{eqnarray}
\iota_0(z)=& -i \log \left\{ \frac{3 C(z) \left(2+C(z)\right)^2}{\left(1+2 C(z)\right)^3} \right. \times \nonumber\\
& \times \left. \left[ 1 + \tanh \left( \frac{1}{2} \log \left[ \frac{ 1 + 2 C(z)}{3}  \right] \right) \right] \right\} - N(z)
\end{eqnarray}
where we have used the auxiliary function, 
\begin{eqnarray}
N(z)= \frac{3}{2} z (2 A+B z)-\arg\left[C(z)\right].
\end{eqnarray}
Note that $N(z)$ was chosen unequivocally such that $y_0(z)$ and $\iota_0(z)$ are continuous functions while equations Eq.(\ref{eq:mu:min}) to Eq.(\ref{eq:iota:min}) remain 
unaffected. 
Figure \ref{fig:Fig6} shows the propagation trajectories of two random initial fields impinging at an equilateral coupler where all waveguides are getting closer together producing an effective coupling parameter that depends linearly on the propagation distance, $g(z) = 1 + 0.5 z$.
Note that such an array protects the eigenstates of the initial configuration through propagation.

\section{Beyond the standard approach}

Finally, we want to show that a theoretical physics treatment of photonic lattices can go beyond the standard approach to the quantum mechanical analogy that considers an array of coupled waveguides as an array of coupled resonators, for the triangular waveguide coupler,
\begin{eqnarray}\label{eq:qmHamiltonian}
	\hat{H}(t) = \sum_{j=1}^{3} \omega_j(t) \hat{a}^{\dagger}_j \hat{a}_{j} + \sum_{j\neq k=1}^{3} g_{jk}(t) \hat{a}^{\dagger}_j \hat{a}_k,
\end{eqnarray}
where the operators $\hat{a}_{j}$ ($\hat{a}_{j}^{\dagger}$) annihilate (create) a photon in the $j$th resonator, and realizes that, with the single photon limit ansatz,
\begin{eqnarray}\label{eq:sglPhoton}
	\left\vert \psi(t) \right\rangle = \mathcal{E}_{1}(t) \left\vert 1, 0, 0 \right\rangle  +
	\mathcal{E}_{2}(t) \left\vert 0,1,0 \right\rangle + \mathcal{E}_{3}(t) \left\vert 0,0,1 \right\rangle, \nonumber \\
\end{eqnarray}
with $ \sum_{j} \vert \mathcal{E}_{j}(z) \vert^2 = 1$, it yields the same system of differential equations,
\begin{eqnarray} \label{eq:}
	i \partial_t	
	\left( \begin{array}{c} 
		\mathcal{E}_{1}(t) \\
		\mathcal{E}_{2}(t) \\
		\mathcal{E}_{3}(t)
	\end{array} \right) =
	\left( \begin{array}{ccc} 
		\omega_{1}(t)  & g_{12}(t) & g_{13}(t) \\
		g_{12}(t) & \omega_{2}(t) & g_{23}(t) \\
		g_{13}(t) & g_{23}(t) & \omega_{3}(t)
	\end{array} \right)
	\left( \begin{array}{c} 
		\mathcal{E}_{1}(t) \\
		\mathcal{E}_{2}(t) \\
		\mathcal{E}_{3}(t)
	\end{array} \right),
	 \nonumber \\
\end{eqnarray}
 as coupled-mode theory, Eq.(\ref{eq:CMT}), once we make the substitution $t=-z$. 
This analogy allowed us to pursuit a group theory approach in the previous sections for the propagation of the field amplitudes but it can also allow us to go further and derive equations of motion for different sets of variables.

\subsection{Classical limit and intensity-phase equations}
For example, if we forget about the single-photon limit and take the classical limit, where the states of light can be taken as coherent states with a high number of photons, then, the canonical pair provided by the creation and annihilation operators can be replaced by the canonical pair of classical intensity, $n_{j}$, and phase, $\phi_{j}$, such that $\hat{a}_{j} \rightarrow \sqrt{n_{j}} e^{i \phi_{j}}$. As a result, we obtain a classical Hamiltonian, 
\begin{eqnarray}\label{eq:clHamiltonian}
	H(z) = \sum_{j=1}^{3} \omega_{j}(z) n_{j}  
	+ \sum_{j \neq k = 1}^{3} g_{jk}(z) \sqrt{n_{j} n_{k}} \cos \left( \phi_{j} - \phi_{k} \right), \nonumber \\
\end{eqnarray}
that provides us with the following equations of motion for the classical canonical pairs, 
\begin{eqnarray}
	\partial_z n_j(z) &=& -\frac{\partial H(z)}{\partial \phi_j}\\
	\partial_z \phi_j(z) &=& \frac{\partial H(z)}{\partial n_j},
\end{eqnarray}
where we have already accounted for the switch from time to propagation distance.
These six coupled differential equations for the three degrees of freedom can be reduced to two degrees of freedom if we substitute the intensity $n_{3}(z) = 1 - n_{1}(z)- n_{2}(z)$ and use the phase $\phi_{3}(z)$ as reference, such that we define the phase differences $\delta_{j}(z) = \phi_{j}(z) - \phi_{3}(z)$ for $j=1,2$, ending up with just four coupled differential equations.
This result is the coupled mode theory description of the propagation of intensity and phase.

\subsection{Classical mechanics and propagation envelopes }
\begin{figure}[htbp]
	\centering  \includegraphics[scale= 1]{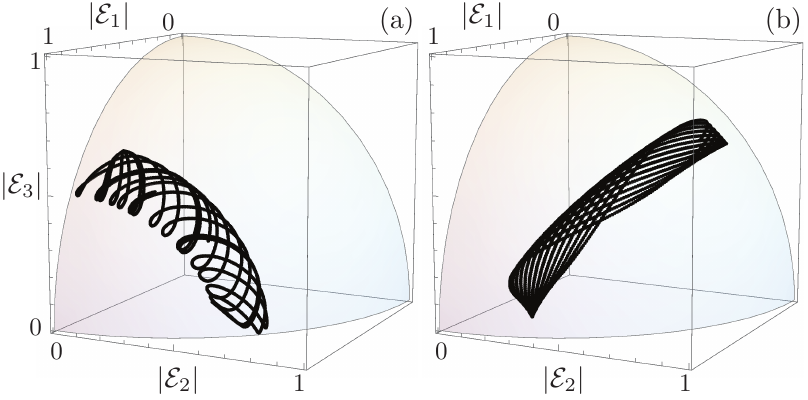}
	\caption{(Color online) Propagation trajectories of absolute amplitudes, $(\vert \mathcal{E}_{1}(z) \vert, \vert \mathcal{E}_{2}(z) \vert, \vert \mathcal{E}_{3}(z) \vert )$, for initial random fields impinging random triangular three-waveguide couplers with constant parameters yielding ergodic trajectories.}
	\label{fig:Fig7}
\end{figure}

We can also choose the standard classical mechanics canonical pair  provided by the field quadratures, 
\begin{eqnarray}
	\hat{q} = \frac{1}{\sqrt{2}} \left( \hat{a}^{\dagger} + \hat{a} \right), ~ 
	\hat{p} = \frac{i}{\sqrt{2}} \left( \hat{a}^{\dagger} + \hat{a} \right),
\end{eqnarray}
and, again, consider classical variables to obtain equations of motion for the field quadratures,
\begin{eqnarray}
	i \partial_z \vert p(z) \rangle	
	 &=&
	H(z) \vert q(z) \rangle , \\
-	i \partial_z \vert q(z) \rangle	 &=& 
    H(z)	\vert p(z) \rangle,
\end{eqnarray}
where, again, we have used kets to represent column vectors, $\vert x(z) \rangle = \left( x_{1}(z), x_{2}(z), x_{3}(z) \right)^{T}$ with $x=p,q$, and made the change from time to propagation distance.
Before moving on, let us focus on the absolute value of the field amplitudes that we have been using to plot the dynamics of our optical trimers, $(\vert \mathcal{E}_{1} \vert,\vert \mathcal{E}_{2} \vert,\vert \mathcal{E}_{3} \vert )$.
These square roots of the intensities at the three waveguides can be understood as a Poincar\'e phase space in the classical mechanics sense.
Then, we can follow a standard classical mechanics approach to find the action-phase variables of the system \cite{Goldstein1980}. 
If the phases are commensurate, the trajectories in phase space will be periodic, and if the phases are incommensurate, the trajectories will be ergodic and fill a region of the phase space defined by the energy of the initial state,
\begin{eqnarray}
\epsilon = \langle \mathcal{E}(0) \vert \hat{H}(0) \vert \mathcal{E}(0) \rangle.
\end{eqnarray}  
For mode-coupling matrices that do not depend on the propagation distance, if the eigenvalues are commensurable, the trajectories in phase space are well-defined and closed, otherwise they will fill a region defined by the initial energy.
Figure \ref{fig:Fig7} shows propagation trajectories of the absolute field amplitudes for the numerical propagation of random parameter sets and initial conditions yielding incommensurate eigenvalues and, thus, ergodic trajectories.

\begin{figure}[htbp]
	\centering \includegraphics[scale= 1]{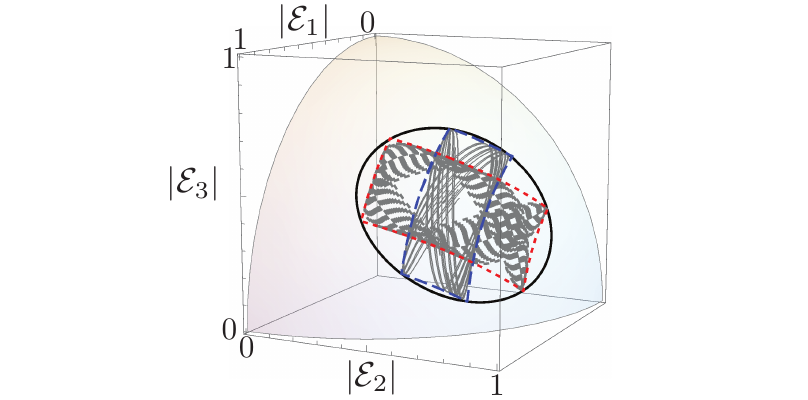}
	\caption{(Color online) Propagation trajectories of absolute amplitudes, $(\vert \mathcal{E}_{1}(z) \vert, \vert \mathcal{E}_{2}(z) \vert, \vert \mathcal{E}_{3}(z) \vert )$,  for two 
		distinct trajectories pertaining to the energy $\epsilon$ (solid gray). The boundary of all possible trajectories (solid black) and the envelopes of the individual  trajectories (dashed blue and dotted red) are calculated according to equation Eq.(\ref{eq:EnvInd}).} 
	
	\label{fig:Fig8}
\end{figure}

Let us focus our attention now on the vast class of incommensurate trajectories and, for the sake of simplicity, keep using optical devices that do not change with the propagation distance.
In many design circumstances, the exact point-to-point evolution of the field amplitudes is not of primary importance and it suffices to gain information on the general region that will be traversed. 
This is characterized by the boundary which encompasses the very region. 
It proves viable to notice that for a given energy, $\epsilon$, and a given set of parameters, $\omega_{j}$ and $g_{jk}$, an infinity number of distinct trajectories prevails.
The conjunction of all these propagation trajectories pertaining to a given energy span a path-connected surface.
We will now find the outer boundary, or envelope, of all propagation trajectories related to an initial energy $\epsilon$.
Equation (\ref{eq:clHamiltonian}) readily reveals that the energy is maximized for given intensities $n_1$, $n_2$ and $n_3$ when all three phases are zero. 
Consequently, the boundary is defined by all triplets $\left( n_1,n_2,n_3 \right)$ that fulfill the following condition,
\begin{equation}
	H_{B}(n_1,n_2,n_3) = \epsilon,	
\end{equation}
with 
\begin{eqnarray}
H_{B}(n_1,n_2,n_3) = \sum_{j=1}^{3} \omega_{j} n_{j}  
	+ \sum_{j \neq k = 1}^{3} g_{jk} \sqrt{n_{j} n_{k}}.
\end{eqnarray}
Due to our normalization requirements, $\sum_{j=1}^{3} n_{j} = 1$, the three intensities are not independent and we can safely omit the third intensity in the following discussion,  
$H_{B}(n_1,n_2,n_3) \rightarrow H_{B}(n_1,n_2)$.
Numerically, once a given set of initial intensities or energy is chosen, the envelope is characterized by the following differential equation,
\begin{eqnarray}
	\frac{\partial H_{B}}{\partial n_2}\delta n_2 + 
	\frac{\partial H_{B}}{\partial n_3}\delta n_3 = 0.
\end{eqnarray}
Figure \ref{fig:Fig8} shows this envelope for all possible trajectories with a given energy $\epsilon$ as a solid black line.

Now, we can turn to calculating the envelope of a single trajectory. 
We have already discussed how to find the propagator for any given three-waveguide coupler. 
In the particular case at hand, constant effective refractive indices and couplings, the propagated field can be written as, 
\begin{eqnarray}
	\vert \mathcal{E}(z) \rangle =  \sum_{j=1}^{3} c_j e^{i \gamma_j z} \vert e_j \rangle,
\end{eqnarray}
where the vectors $\vert e_j \rangle$ are the three normal modes of the coupler with corresponding eigenvalues $\omega_j$ and the normal mode decomposition parameters are defined as the inner product between the normal mode and the initial condition,  $c_{j} = \langle e_{j} \vert \mathcal{E}(0) \rangle$. 
We can introduce a phase relevant parameter that distorts the ratio between eigenvalues, 
\begin{eqnarray}
\vert \mathcal{E}(z, \phi) \rangle = c_1 e^{i \left( \gamma_1 z + \phi \right) } \vert e_1 \rangle + c_2 e^{i \gamma_2 z} \vert e_2 \rangle + c_3 e^{i \gamma_3 z} \vert e_3 \rangle,
 \nonumber \\
\end{eqnarray}
The parameter $\phi$ offsets a given trajectory. 
However, this cannot hold true at the envelope and, hence, we can establish the following condition for the envelope,
\begin{eqnarray}
	\frac{ (\frac{\partial n_1}{\partial z}) }{ (\frac{\partial n_2}{\partial z}) } = \frac{( \frac{\partial n_1}{\partial \phi})}{(\frac{\partial n_2}{\partial \phi}) } ,
\end{eqnarray}
which is identical to requiring a null determinant,
\begin{eqnarray}
	D(z, \phi) &=& \left\vert \left(  \begin{array}{cc} 
		\frac{\partial n_1}{\partial z} & \frac{\partial n_2}{\partial z} \\
		\frac{\partial n_1}{\partial \phi} & \frac{\partial n_2}{\partial \phi} 
	\end{array} \right) \right\vert, \\
	&=& 0.
\end{eqnarray}
Accordingly, the envelope can be calculated from the following differential equation,
\begin{eqnarray}
	\frac{\partial D(z,\phi)}{\partial z} \delta z +
	\frac{\partial D(z,\phi)}{\partial \phi} \delta \phi = 0. \label{eq:EnvInd}
\end{eqnarray}
Figure \ref{fig:Fig8} shows two examples of envelopes for individual trajectories with particular initial field amplitudes and the same energy $\epsilon$ as dashed blue and dotted red lines. 

\section{Conclusion}
We have shown that it is possible to approach the propagation of electromagnetic field through triangular three-waveguide couplers, in terms of the Lie algebra $su(3)$. 
We derived a formal solution \`a la Wei-Norman in line with the idea of Gilmore-Perelomov coherent states.
In order to provide practical examples, we focused our attention on a reduced class of structures where the waveguides are identical and the effective couplings are constant, independent of the propagation distance, and christened them constant optical trimers. 
We established a connection between the equilateral trimer, actually any circular array of waveguides with identical couplings, and the discrete Fourier transform, and between the isoceles trimer and the golden ratio where it is possible to suppress the energy transfer to one of the waveguides. 
For the sake of providing a propagation dependent example, we studied a system where all the waveguides get symmetrically closer or further apart following a linear z-dependent effective coupling.
Our results for passive triangular three-waveguide couplers may be useful for a theoretical approach to linear and nonlinear active devices.

We went beyond the standard classical-quantum analogy in the single photon limit and showed that, using the classical limit, it is possible to recover the coupled mode differential equations describing the propagation of intensity and phase.
Furthermore, we also showed that working with the classical mechanics canonical pair, which are the equivalent of the electromagnetic field quadratures, allows us to calculate the envelope of all possible trajectories with a given initial energy as well as individual envelopes for specific input field amplitude distributions.
All these results may provide a working platform for the design of photonic integrated devices where the main interest is to generate a specific output from a set of particular inputs instead of studying point-to-point propagation through the devices.

We hope this manuscript could also serve as a tutorial for the use of theoretical physics tools in the design of photonic lattices and, viceversa, encourage theoretical physicists to make incursions on the field of photonic circuit design.





\section*{References}

\end{document}